\documentclass[11pt]{article}
\usepackage{moriond,epsfig}

\usepackage{amsmath}
\usepackage{amssymb}
\usepackage{epsfig}
\usepackage{graphicx}
\usepackage{cite}
\usepackage{multirow}
\usepackage{longtable}
\usepackage{lscape}
\usepackage{bbm}
\usepackage{dcolumn}
\usepackage[dvips]{color}
\usepackage{bm}
\usepackage{booktabs}
\usepackage{wick}
\usepackage{rotating}
\usepackage{mathrsfs}

\usepackage{pstricks}
\usepackage{pst-node}
\usepackage{feynmp}
\usepackage{slashed}
\usepackage{latexsym}

\newcommand{\bi}{\begin{itemize}}
\newcommand{\ei}{\end{itemize}}

\newcommand{\be}{\begin{equation}}
\newcommand{\ee}{\end{equation}}
\newcommand{\bea}{\begin{eqnarray}}
\newcommand{\eea}{\end{eqnarray}}
\newcommand{\nn}{\nonumber}

\newcommand{\rw}{\rightarrow}

\newcommand{\mcl}[1]{\mathcal{#1}}
\newcommand{\trm}[1]{\textrm{#1}}
\newcommand{\msc}[1]{\mathscr{#1}}
\newcommand{\capdef}{}
\newcommand{\mycaption}[2][\capdef]{\renewcommand{\capdef}{#2}%
        \caption[#1]{{\footnotesize #2}}}

\newcommand{\ie}{{\it i.e.}}

\newcommand{\onsi}{\mathcal{O}_{\mathrm{NSI}}}

\begin{document}
\vspace*{4cm}
\title{Gauge invariant non-standard neutrino interactions}

\author{ D. Hernandez}

\address{Departamento de F{\'i}sica Te{\'o}rica and Instituto de F{\'i}sica Te{\'o}rica UAM/CSIC, \\
        Universidad Aut{\'o}noma de Madrid, 28049 Cantoblanco, Madrid, Spain}

\maketitle\abstracts{Theories beyond the Standard Model must respect its gauge symmetry. This implies strict constraints on the possible models of Non-Standard Neutrino Interactions (NSIs). We review here the present status of NSIs from the point of view of effective field theory. Our recent work on the restrictions implied by Standard Model gauge invariance is provided along with some examples of possible gauge invariant models featuring non-standard interactions.}


\section{Introduction}


The experimental observation of neutrino masses and mixings is the first evidence for physics beyond the Standard Model (SM) -- maybe together with the indication for dark matter -- and points to the existence of a new, yet unknown, physics scale.  The tiny masses of the neutrinos, which are orders of magnitude lighter than those of other fermions, suggest a large new physics scale leading to very suppressed effects.
Since neutrinos have only weak interactions with the SM particles, they may even
constitute an excellent window into the new physics underlying the
``dark sectors'' of the universe, \ie, dark matter and dark energy.
Therefore,  new physics may very well appear next in the form of exotic couplings involving neutrinos, which are often called {\it non-standard neutrino interactions}(NSIs). In the present work this possibility is discussed  in a model independent way and we focus, in particular, in the possible tree-level mediators of new physics inducing them.

On general grounds, whatever the nature of the new couplings is, observable effects will only be expected for a new physics scale $\Lambda$ near the present experimental limits, \ie, above the electroweak symmetry breaking (EWSB) scale.
An example of NSI is given by  the dimension six ($d=6$) operator in
\begin{equation}
\frac{1}{\Lambda^2} (\bar{\nu}_\alpha \gamma^\rho P_L \nu_\beta)\,( \bar\ell_\gamma \gamma_\rho \ell_\delta )\, .
\label{equ:raw}
\end{equation}
where spinor indices have been omitted and flavour indices are represented by greek characters.

While the high energy theory has to contain and encompass the SM gauge group, the operator in Eq.\eqref{equ:raw} is not gauge invariant. A gauge invariant operator that produces this coupling is, for instance, the following
\begin{equation}
\frac{1}{\Lambda^2} (\bar{L}_\alpha \gamma^\rho L_\beta)\,( \bar{L}_\gamma \gamma_\rho L_\delta \,) \, ,
\label{equ:dressed}
\end{equation}
where $L$ denotes the leptonic $SU(2)$ doublet. However, Eq.\eqref{equ:dressed} is a good example of the effects involved in taking gauge invariance into account. The operator above, not only produces the desired combination, Eq.\eqref{equ:raw} but also contributes to the transition $\mu \rw 3e$, for $\beta = \mu$, $\alpha=\gamma=\delta=e$, with the same strength. Hence, the extremely strong bounds for the charged leptons transition also apply to the NSI.

In general, effective interactions such as those in Eq.\eqref{equ:raw}, clearly require to consider operators made out of {\it four} leptonic fields, plus Higgs fields in the case of operators with $d> 6$~\cite{Davidson:2003ha,Bergmann:1998ft,Bergmann:1999pk}. There is a plethora of $d=6$~\cite{Buchmuller:1985jz} and $d=8$~\cite{Berezhiani:2001rs} operators and different classes of models result in different sets of operators and operator coefficients. Those among them relevant for NSI can affect neutrino production or detection processes, or modify the matter effects in their propagation, depending on the operator or combination of operators considered.

The motivation of the present work is to determine what is the minimum complexity needed to achieve large NSIs in the sense of relevant for present or near future experiment. We find that, in all generality, fine tuning is needed to suppress the interactions among four charged leptons. We further provide a tree level classification of all possible models leading to $d=6$ or $d=8$ NSIs. These models, as will be seen, suffer in general, not only from the restrictions mentioned above but also from new ones particular to each. We illustrate this case with one example.


\section{Formalism of effective operators}


From the point of view of effective field theory, the NSI Lagrangian relevant to us can be written as
\begin{equation}
\mathcal{\delta} \mathscr{L }_{\text{eff}} = \frac{1}{\Lambda^2}\sum^{d=6}_i \mathcal{C}_{i} \, \mathcal{O}_i^{d=6} + \frac{1}{\Lambda^4}\sum^{d=8}_k \mathcal{C}_{k} \, \mathcal{O}_k^{d=8}\,,
\label{equ:effop}
 \end{equation}
where the two terms run over all possible $d=6$ and $d=8$ operators relevant for purely leptonic NSIs. The field content of these operators is fixed, namely: the $ \mathcal{O}_i^{d=6}$ are four fermion operators; in the case of $d=8$, the $\mathcal{O}_k^{d=8}$ consist of four fermions plus two Higgses.

In both these series there are certain operators that are not afflicted by the appearance of four charged lepton interactions. It is the case of the gauge invariant $d=6$
\be
\mcl{O}_{\trm{Zee}}^{d=6} = (\bar{L}i\tau_2L)(\bar{L}i\tau_2L) \label{zee}
\ee
which is characteristic of a class of models commonly referred in the literature as Zee models. Also, after EW symmetry breaking, for $d=8$ we have
\be
\mcl{O}_{NSI}^{d=8} = (\bar{L}H)\gamma^\mu(H^\dagger L)(\bar{E}\gamma_\mu E) \label{O_NSI} \,.
\ee
For these two, expansion on the flavour indices show that only interactions involving neutrinos appear. Notice the importance of the Higgs in $\mcl{O}_{NSI}^{d=8}$ at selecting the neutrinos from the lepton doublets \cite{Davidson:2003ha}.

In general however, the operators in Eq.\eqref{equ:effop} will indeed generate four-charged lepton interactions. A key observation is that every operator in these sums can be expressed as a linear combination of a selected few \cite{Buchmuller:1985jz,Berezhiani:2001rs}. Thus, for instance, every $d=6$ operator involving four lepton doublets can be written as a linear combination of the following two
\begin{align}
  (\mathcal{O}_{LL}^{{\bf 1}})^{\beta \delta}_{\alpha \gamma} = &  (\bar{L}^{\beta} \gamma^{\rho} L_{\alpha}) (\bar{L}^{\delta} \gamma_{\rho} L_{\gamma}) \, , \label{equ:oll1} \\ 
  (\mathcal{O}_{LL}^{{\bf 3}})^{\beta \delta}_{\alpha \gamma} = & (\bar{L}^{\beta} \gamma^{\rho} \vec{\tau} L_{\alpha}) (\bar{L}^{\delta} \gamma_{\rho} \vec{\tau} L_{\gamma})
\end{align}
Both these operators yield charged lepton couplings. However, if both are present in the effective Lagrangian, their coefficients $\mathcal{C}_{LL}^{{\bf 1}}$, $\mathcal{C}_{LL}^{{\bf 3}}$ can be 'tuned' in such a way that the charged interactions are suppressed leaving only large NSIs. The condition for this example is simply
\be
\mathcal{C}_{LL}^{{\bf 1}} = - \mathcal{C}_{LL}^{{\bf 3}} \, .
\ee

In \cite{Gavela:2008ra} we exhausted all the cancellation relations for $d=6$ and $d=8$ operators. We found that this kind of cancellation conditions are necessary for each model that attempts to produce large NSIs. The operators $\mcl{O}_{\trm{Zee}}^{d=6}$ or $\mcl{O}_{NSI}^{d=8}$ are clearly particular cases where such conditions are fulfilled automatically.

Nevertheless, in general, the coefficients $\mathcal{C}_{LL}^{{\bf 1}}$, $\mathcal{C}_{LL}^{{\bf 3}}$, etc. will be given in terms of the parameters of the fundamental Lagrangian. Stepping into the model building arena, one is tempted to ask: how are these cancellations implemented in terms of the fundamental couplings of a particular model?


\section{A toy model}


 In order to estimate the theoretical price to pay for obtaining large NSIs  without large charged lepton flavour violation, we show here a toy model in a bottom-up fashion, which precisely generates the $d=8$ operator $\onsi$ in Eq.~(\ref{O_NSI}) and no $d=6$ operator. Then we will sketch a systematic analysis, from which we can recover the toy model as the simplest possibility in a top-down approach.

Consider the following  Lagrangian for the underlying theory, which adds both a new scalar doublet $\Phi$ and a vector doublet $V_\mu$  to the SM Lagrangian, with general couplings  to the SM fields $y$, $g$ and $\lambda$'s,
 \begin{align}
\mathscr{L} & = \mathscr{L}_{\text{SM}}
  - {(y)_{\beta}}^{\gamma}\, (\bar{L}^\beta)^{i} E_\gamma \Phi_{i}
  - (g)_{\beta \delta} \, (\bar{L}^\beta)^{i}  \gamma^{\rho}
  (E^c)^\delta (V_\rho)_{i}
  \nn \\ &
  \quad
  + \lambda_{{\bf 1}s}(H^\dagger H)(\Phi^\dagger \Phi)
  + \lambda_{{\bf 3}s}(H^\dagger \vec\tau H)(\Phi^\dagger \vec\tau \Phi)
  \nn \\
  & \quad
  + \lambda_{{\bf 1}v}(H^\dagger H) (V_\rho^\dagger V^\rho)
  + \lambda_{{\bf 3}v}(H^\dagger \vec\tau H)(V_\rho^\dagger \vec\tau
  V^\rho)
  + \text{h.c.} \,+...
  \label{equ:toy}
\end{align}
where the dots refer to other bosonic interactions not relevant for this work. This model leads both to $d=6$ and $d=8$ charged interactions as well as NSIs at $d=8$. In Fig.(\ref{fig:model1}) we include the two diagrams contributing to the $d=8$ NSI.
\begin{figure}[t]
\begin{center}
\unitlength=1.0cm
\begin{picture}(8,6)
  \includegraphics{LLLLHH-scalar.1}
  \hspace*{0.8cm}
  \includegraphics{LLLLHH-vector.1}
\thicklines
\put(-5.2,0.3){$E$}
\put(-5.2,5.1){$L$}
\put(-8.0,0.3){$L$}
\put(-8.0,5.1){$E$}
\put(-5.2,3.0){$H$}
\put(-8.0,3.0){$H$}

\put(-0.6,0.3){$E$}
\put(-0.6,5.1){$E$}
\put(-3.4,0.3){$L$}
\put(-3.4,5.1){$L$}
\put(-0.6,3.0){$H$}
\put(-3.4,3.0){$H$}
\end{picture}
\end{center}
\mycaption{\label{fig:model1} Dimension eight operator for the toy model decomposed into dimension four interactions. The mediator particles are respectively a scalar doublet and a vector doublet of the SM $SU(2)$}
\end{figure}
As we have stated, in order to suppress the four charged interactions we need to impose certain relations between the couplings in the Lagrangian Eq.\eqref{equ:toy}. It can be shown \cite{Gavela:2008ra} that the cancellation condition that eliminates the $d=6$ charged interactions is given by
\be \label{equ:cond-1-for-Yukawas}
-\,2({g}^\dagger)^{\gamma \alpha} (g)_{\beta\delta} + {({y}^\dagger)_{\delta}}^{\alpha} {(y)_{\beta}}^{\gamma} = 0\,.
\ee
while, in order to cancel the $d=8$ charged interactions, we need to impose additionally that
\be
\lambda_{{\bf 1} s} + \lambda_{{\bf 1}v} = \lambda_{{\bf 3} s} + \lambda_{{\bf 3} v} \neq 0\,.
\label{equ:lambda}
\ee
It can now be proven that, with these conditions, amplitudes for NSI processes remain large. It is beyond the purpose of this work to do a phenomenological analysis of this model. What is important is to stress that a toy model for viable NSI has resulted by adding a pair of exotic particles to the SM field content \emph{and} imposing the two relations to their couplings with the SM fields contained in Eq.\eqref{equ:cond-1-for-Yukawas} and \eqref{equ:lambda}. Unfortunately, it also illustrates how these ad hoc cancellations are required a priori. It remains a question for the model builder whether they can be justified by some some symmetry or not.


\section{General model analysis}


If one restricts oneself to the tree level analysis, then there is a finite number of models that can generate any effective operator. In the case of the $d=6$ operators the models that correspond to each can be quickly identified, since there is only one possible type of tree level diagram that generates such effective couplings. 

As an illustration, consider the following mediator decomposition of operator $\mcl{O}_{\trm{Zee}}^{d=6}$ of Eq.\eqref{zee}. 

\vspace{.4cm}
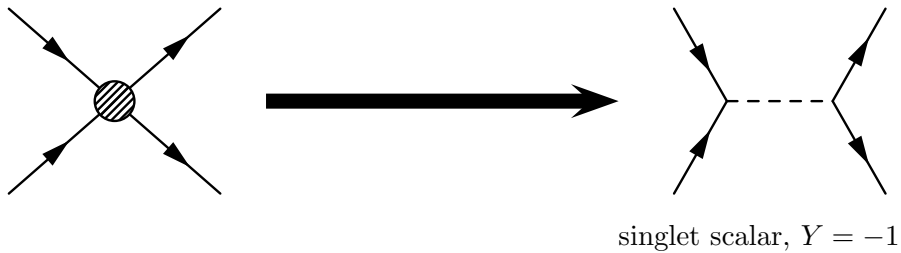
\begin{center}
  \begin{tabular}[h]{p{5cm} p{3cm} p{5cm}}

    \rnode{A}{
      \begin{fmffile}{diag5}
        \begin{fmfgraph*}(100,70)
          \fmfleft{i1,i2}
          \fmfright{o1,o2}
          \fmf{fermion,label={\small $L^e$}}{i1,v1}
          \fmf{fermion,label={\small $L^e$}}{i2,v1}
          \fmf{fermion,label={\small $L^e$},label.side=right}{v1,o1}
          \fmf{fermion,label={\small $L^\mu$},label.side=left}{v1,o2}
          \fmfblob{.15w}{v1}
        \end{fmfgraph*}
      \end{fmffile}
    } 
    
    &
    \hspace{2cm}
    &
    %

     \rnode{B}{
       \begin{fmffile}{diag6}
        \begin{fmfgraph*}(100,70)
           \fmfleft{i1,i2}
           \fmfright{o1,o2}
           \fmf{fermion,label={\small $L^e$}}{i1,v1}
           \fmf{fermion,label={\small $L^e$}}{i2,v1}
           \fmf{dashes,label={\small \rnode{S}{$S$} } }{v1,v2}
           \fmf{fermion,label={\small $L^e$},label.side=left}{v2,o1}
           \fmf{fermion,label={\small $L^\mu$},label.side=right}{v2,o2}
         \end{fmfgraph*}
       \end{fmffile}
     }
     \ncline[linewidth=.2cm]{->}{A}{B}

     \vspace{.2cm}
     \rnode{T}{singlet scalar, $Y=-1$} 

  \end{tabular}
  \ncline[nodesep=4pt]{->}{S}{T}
\end{center}

\vspace{-.2cm}
\noindent In the figure it is illustrated the procedure to obtain the unique model that generates it at tree level. One begins by grouping the operator into bilinears and identifying those as external legs from a common vertex. Then the particle that connects those vertices can be uniquely determined.

It is important to take notice at this point that, once a model is chosen, the general constraints involving some kind of 'tuning' that we saw earlier are usually accompanied by new ones, typical of that particular model. Strong bounds for the model above, characterized by additional interaction terms in the Lagrangian of the form
\be
\delta \msc{L}_{int} = S\bar{L}i\tau_2 L + \trm{h.c.}
\ee
have been deduced in \cite{Antusch:2008tz}. It is not hard to prove that essentially all NSI from $d=6$ operators are strongly constrained when the possible tree-level mediators are taken into account \cite{Gavela:2008ra} .

The $d=8$ case is more involved but one can analyze it following essentially the same procedure. As we saw in the toy model, apart from the usual cancellations required to suppress the four-charged processes, in general one might have to impose additional cancellation conditions in order to ensure that no $d=6$ operator is generated. In \cite{Gavela:2008ra} a systematic analysis of all the cancellation conditions for $d=8$ operators, along the lines that have been presented for the $d=6$ case, was carried out. It was shown there that at least two new fields are required to avoid the undesired $d=6$ and $d=8$ interactions involving four charged leptons  In fact, when the mediators of a $d=8$ effective operator couple only to SM bilinears, 
there will always be at least one field leading as well to $d=6$ contributions. These have to be cancelled in each case by fine-tuning or symmetries.


\section{Conclusions}


In this talk, we have discussed the possibility of large non-standard interactions (NSI) in the neutrino sector. Since any model of new physics has to recover the Standard Model at low energies, we have required gauge invariance under the SM gauge group and studied the possible effective theories. The focus is set on  purely leptonic NSI, that is,  on operators in which the only fermion fields appearing are leptons. Our analysis is based on the full (analytical) decomposition of possible dimension six and eight effective operators, which can be induced at tree-level by any hypothetical beyond the SM theory. 

The aim is to gauge the theoretical price of achieving phenomenologically viable large neutrino NSI. It is argued that the minimum complexity of a realistic model leading to large NSI and no charged lepton flavor violation requires at least two new fields inducing $d=8$ NSI couplings. Furthermore, the possible SM charges of those mediators and the cancellation conditions for the dimension six  interactions among four leptons that they simultaneously induce can also be found \cite{Gavela:2008ra}. These cancellation conditions translate into precise relations among model parameters. We have shown how this works in a particular case. Our results imply nonetheless such a number of constraints that the observational prospects do not seem bright, specially as we did not identify some symmetry which would account for them. On the other side, we showed that large NSI are not excluded, and we found out which conditions are necessary to satisfy for any model to be viable. For the model to be credible, those cancellation conditions should be explained by some symmetry. Until then, it's up to the reader to decide on the perspective for large NSI.


\section*{Acknowledgements}


This work has been done in collaboration with M.B. Gavela, T. Ota and W.Winter and the author was partially supported by CICYT through the project FPA2006-05423 as well as from the Comunidad Aut\'{o}noma de Madrid through Proyecto HEPHACOS. The author also acknowledges support from the MEC through FP grant AP20053603 as well as the LPT (Orsay) for hospitality during the last stage of this work.

\end{document}